\documentclass[twocolumn,showpacs,preprintnumbers,amsmath,amssymb]{revtex4}
\usepackage{graphicx}% Include figure files
\usepackage{dcolumn}% Align table columns on decimal point
\usepackage{bm}% bold math
\usepackage{color}% color

\begin{document}
\title{Buckled in translation}
\author{E. Wandersman\footnote{present adress: Kamerlingh Onnes Lab, Universiteit Leiden, Postbus 9504, 2300 RA, Leiden, The Netherlands}}
\author{ N. Quennouz}%
\author{M. Fermigier}%
\author{A. Lindner\footnote{corresponding author : anke.lindner@espci.fr}}%
\author{O. du Roure}%
\affiliation{Laboratoire de Physique et M\'ecanique des Milieux
H\'et\'erogenes- UMR 7636 CNRS/ESCPI Paristech  - Universit\'e
Pierre et Marie Curie - Universite Paris Diderot - 10, rue Vauquelin
- 75005 Paris -France}
\begin{abstract}
     We report experiments on the deformation and transport of an elastic fiber in a viscous cellular flow, namely a lattice of counter-rotative vortices. We
show that the fiber can buckle when approaching a
stagnation point. By tuning either the flow or fiber
properties, we measure the onset of this buckling instability. The buckling threshold is determined by the relative intensity of viscous and elastic forces, the elasto-viscous number $S_p$. Moreover we show that
flexible fibers escape faster from a vortex (formed by
closed streamlines) compared to rigid fibers. As a
consequence, the deformation of the fiber changes its
transport properties in the cellular flow.
 \end{abstract}
\maketitle

\par

The interaction of a deformable body with a viscous flow
is found in a wide range of situations, ranking from biology to
polymer science. Cells or micro-organisms in water use deformation
of high-aspect ratio flexible organelles called cilia or flagella to
move at low Reynolds number \cite{Bray}. Numerous studies have been
dedicated to locomotion and hydrodynamic interactions at low
Reynolds number, including the study of biological
\cite{GollubScience2009, Abkarian2007} or model systems
\cite{Nais1,qian2008,Peko2006,DreyfusNat06, Abkarian2007,Aranson}, but a large number
of questions remains still to be
answered. The stretching of a flexible polymer in a viscous flow is
well studied \cite{Perkins}. The so called
coil-stretch transition is responsible for the
spectacular macroscopic properties of polymer solutions as for
example normal stresses or high elongational viscosities.
This transition might have an analog for elastic fibers that buckle
by viscous forces. Microscopic buckling of
fibers has indeed been shown theoretically to lead to original rheological properties of their
suspensions, such as the appearance of normal stress differences \cite{BeckerPRL01}. An
other fundamental question is the modification of the translational
dynamics of an object induced by its deformation in a viscous flow
as studied by Young and Shelley \cite{ShelleyPRL07}.

\begin{figure}[!ht]
\begin{center}
\includegraphics[width=7cm]{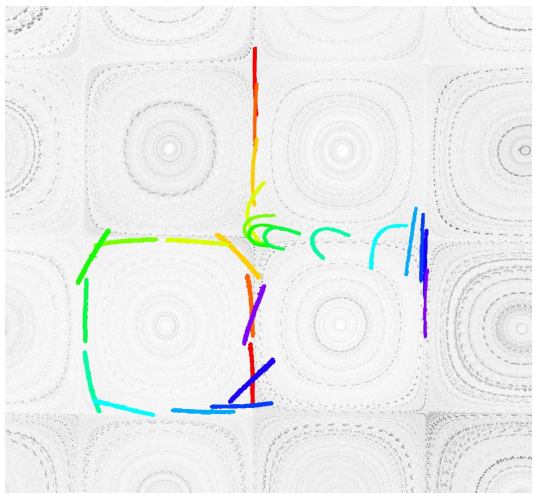}
\caption{\label{fig1} %Typical trajectories
Successive shapes of two fibers experiencing the same flow field but with different Young moduli ($\Delta t=0.4 s$, from red to purple). The more rigid fiber (bottom left) does not buckle whereas buckling is observed for the more flexible fiber (up right). The background flow is visualized
independently.}
\end{center}
\end{figure}

The complex dynamics of such systems originates from different
sources: first, the dynamics is dependent on the size of the
deformable body with respect to the characteristic lengthscales of
the flow. Second, when the structure is deformed by viscous forces, the
local surrounding flow will be modified leading to non-linear
effects. Hydrodynamic interactions between flexible objects further
increase complexity \cite{Young2009}.
The experimental study of simple model systems, such as
homogeneous elastic fibers, evolving in a controlled flow geometry,
could help understanding these complex interactions.

In this communication we experimentally study the deformation of an
isolated macroscopic filament approaching a stagnation point and the
subsequent modification of its transport through a cellular flow of
counter-rotating vortices (see figure \ref{fig1}). First we
characterize the buckling transition of the fiber varying the
properties of the fiber and the viscous flow independently and
identify the buckling threshold. The threshold is shown to be
function of the relative intensity of the viscous and elastic
forces, the so called $S_p$ number \cite{Wiggins1998}. We then
discuss differences in the transport of rigid and flexible fibers in
the cellular flow and show that deformation significantly changes the
transport properties.

\begin{figure}[t!]
\begin{center}
\includegraphics[width=8cm]{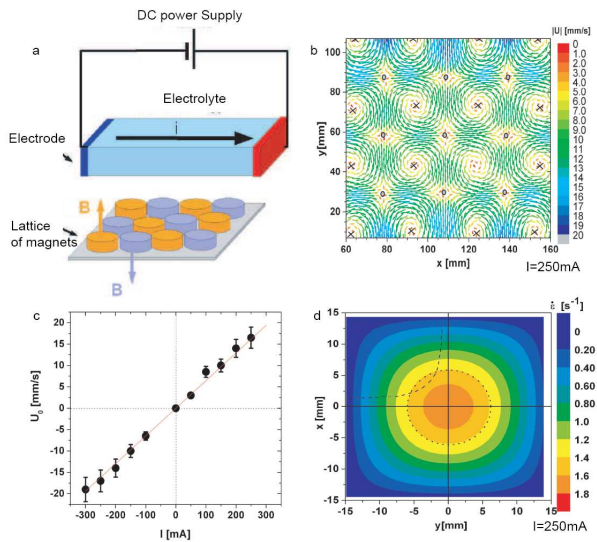}
\caption{\label{fig:setup}  a) Experimental
setup: a lattice of magnet with alternated orientation
is placed below an electrolyte.  b) Velocity
field. Closed circles represent stagnation points; crosses, centers of the vortices. c) Maximum velocity $U_0$ vs electric current $I$. The line is the best linear fit. d) Color mapping of the local compression
rate $\dot{\epsilon}$ in a cell. The dashed line is one experimental trajectory of the fiber. The dotted circle
represents 0.8 $\dot{\epsilon}^{max}$.}
\end{center}
\end{figure}

The lattice of counter-rotating vortices is obtained using
electromagnetic forcing \cite{CardosoPRE95,Twardos2008}. An electric current is applied to
an electrolyte placed in a rectangular cell (Fig. \ref{fig:setup}a).
A lattice of 4x5 magnets (NdFeB, Supermagnets) with alternating orientation is
placed underneath the cell, creating a spatially periodic magnetic
field. Locally varying Lorentz forces act on the flowing electrolyte
leading to an array of counter-rotating vortices of size $W=$ 3~cm. The electrolyte is a 50:50 mixture of
Polyethyleneglycol 1000 (Fluka) and purified water, into which NaCl
is added up to saturation(viscosity, $\eta=35\pm5$ mPa.s, surface tension $\gamma=43\pm1$ mN/m. The electric current $I=-400/+400$ mA is applied via a DC power supply
using carbon fiber electrodes (Toyobo).
\par The array of vortices generates a cellular flow and an array of
stagnation points. Close to the stagnation points the flow can be
described by hyperbolic streamlines, corresponding to a purely
elongational flow with constant elongation/compression rate
$\dot{\epsilon}$. Close to the center of the vortices the flow can
be described by solid rotation without any deformation of the fluid.
Note that the flow consists of closed streamlines and small
particles can not be transported across the flow field when
diffusion and particle inertia are neglected.

The velocity field is determined by Particle Imaging
Velocimetry (using DAVIS software) and presented on figure \ref{fig:setup}.  All our experiments are performed  in a selected region, a 3x3 lattice of stagnation points, where boundary effects are
negligible. The velocity field is well described by the following equation:
\begin{equation}
\vec{U}(\tilde{x},\tilde{y})=U_0 \left(
                               \begin{array}{c}
                                 sin(\tilde{x}) cos(\tilde{y}) \\
                                 - cos(\tilde{x}) sin(\tilde{y}) \\
                               \end{array}
                             \right)
\label{eq:U}
\end{equation}
\noindent where $\tilde{x}$ (resp. $\tilde{y}$) is the dimensionless
coordinate $\pi x/W$ (resp. $\pi y/W$). $\tilde{x}$ and $\tilde{y}$ are set to
zero at a stagnation point. The parameter $U_0$ is found to be proportional to the electric
current $I$ (see fig. \ref{fig:setup}c). Note that in the range of velocities used the
Reynolds number is always below $5$. 
The local compression rate of the flow reads:
\begin{equation}
\label{equ:eps} \dot{\epsilon}(\tilde{x},\tilde{y},I)=\pm \frac{\pi
U_0(I)}{W}cos(\tilde{x})cos(\tilde{y})
\end{equation}
\noindent with opposite signs considering either compressive or
extensive quadrants. The direction of compression remains within one
quadrant always the same. The compression rate is locally
variable, being maximum at the stagnation point
($\dot{\epsilon}_{max}= \pi U_0 /W$), and zero at the center of the
vortices.

During its motion in the lattice, a fiber thus experiences a
variable compression rate $\dot{\epsilon}(\tilde{x},\tilde{y})$.
This is illustrated in figure \ref{fig:setup}d, where the local
compression rate is color mapped in a cell centered on a stagnation
point. All our experimental trajectories reach 80$\%$ of
$\dot{\epsilon}_{max}$. Thus, for sake of
simplicity, we will hereafter use this value to characterize the flow influence.\\

Experiments are performed with macroscopic fibers (length $L$=12mm $\pm$ 0.5  mm,
radius $r$=85 $\pm$ 5  $\mu$m) made by crosslinking vinylpolysiloxane with a curing agent
(Zhermack, 8 Shore A) in a glass capillary. The filament is then carefully extruded from
the capillary. Optical measurements on a microscope have been used
to measure the radius of the fiber as well as to verify that it is homogeneous and not damaged over the whole length. The ratio between its length and the cell
size $\alpha=L/W=0.4$ is fixed for all experiments. The mass ratio between the curing agent and the polymer can be varied from 1:2 to 3:1. The Young's
modulus is obtained via rheological measurements using $Y=3G'$ and can be tuned
from $Y$=75~kPa to $Y=180$~kPa.

The density of the fiber is lower than the density of the fluid ($\Delta\rho=0.15.10^3$~kg/m$^{-3}$) and
the fiber thus floats at the surface. More precisely, the equilibrium position of the fiber at the interface is determined by the interplay between gravity and surface tension \cite{Rapa1976}. For our system, surface tension dominates, leading to the absence of a meniscus. The degree of immersion of the fiber is than solely controlled by the wetting angle between the liquid and the fiber. 
We find this angle to be $90\pm 5 \deg$ and the fiber is approximatively half-immersed.
As a consequence, the motion as well as the deformation of the fiber
are bidimensional, constrained at the surface of the flow. As the interface is not deformed, the influence of surface tension on the dynamics of the filament is negligible.

Pictures of the fiber are taken with a digital camera (PixelLink, 1024x768,
10 fps). Using standard detection procedures (ImageJ software), we measure the position of the
center of mass of the fiber. A fit
of the shape is performed with spline functions
(Numerical Recipes 3.3),
and used to extract the mean curvature of the fiber as a function of time $t$, $C(t)$.

We will now use the experimental set-up to study the buckling
instability of the flexible fibers in the vicinity of a stagnation
point. Figure \ref{fig1} shows two typical trajectories of a fiber with different elastic
moduli, but evolving in the same flow. One observes that the rigid
fiber does not buckle when approaching the stagnation point, the
flexible fiber however is deformed by the flow and shows a buckling
instability.

To characterize this buckling instability in detail we construct the
probability of a fiber to buckle in a given flow. We consider
trajectories of the fiber cell by cell, where a cell is a square of
width $W$, centered on a stagnation point. One trajectory is defined as the path in the vicinity of one stagnation point as illustrated by the dashed line on figure \ref{fig:setup}d. The fiber is
said in "coiled state" if the mean curvature is larger than a given
threshold $C_{coil}=0.125$ mm$^{-1}$, i.e. to a radius of
curvature of approximatively $2L/3$. This corresponds to almost a quarter of a circle. No quantitative modifications of the results are observed if the threshold is slightly tuned. We do not
consider trajectories where the filament enters the cell in a
coiled state. We determined, for each experiment
the ratio between the number of trajectories displaying a coiled
state $N_{coiled}$ and the total number of trajectories $N_{tot}$.
Assembling over 20 experiments, around 7000 trajectories of the
fiber are considered, with a total number of buckling events equal
to $\approx$500.

The fiber is predicted to buckle when viscous forces overcome
elastic forces. Their relative intensity is given by the Sperm
number $S_p$ \cite{Wiggins1998}:
\begin{equation}
\label{equ:Sp} S_p=\frac{\eta\dot{\epsilon}}{Y}\frac{L^4}{r^4}.
\end{equation}

\begin{figure}[ht!]
\begin{center}
\includegraphics[width=7.8cm,height=7cm]{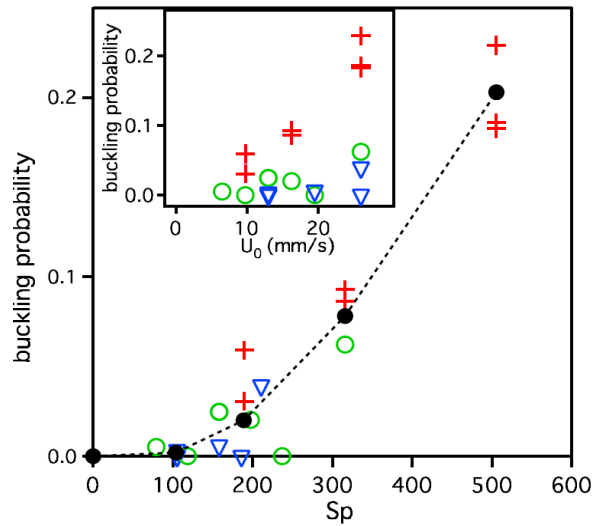}
\caption{\label{fig:proba} Probability of buckling vs
$S_p$ for different Young moduli (red cross Y=75kPa,
blue triangle Y=120 kPa, green open circle Y=180 kPa). Black closed circles represent the mean value. Inset: Probability of buckling as a function of $U_0$.}
\end{center}

\end{figure}

Our set-up allows to vary the properties of the fiber and of the
viscous flow separately tuning the two  forces independently. The inset of figure
\ref{fig:proba} shows the probability to buckle as a function of
the maximum velocity of the flow $U_0$ for different Young's moduli (Y=75, 120 and 180 kPa). One
observes for all cases an increase  with $U_0$. One observes furthermore
that the probability to buckle is higher for the flexible fibers.

Figure \ref{fig:proba} now shows the probability to buckle for all
experiments and its average as a function of $S_p$. Within the
experimental resolution the data points fall on a single curve,
showing that the $S_p$ number is indeed the control parameter of the
system, as predicted by Young {\it et al.} \cite{ShelleyPRL07}. 

The probability to buckle is zero below $S_p=120$, it increases at
$S_p \sim 120-150$ and reaches at our highest $S_p$ about 20$\%$. Theoretical predictions from linear stability
analysis in a hyperbolic flow show a threshold of $S_p^*$=153 for
the buckling instability \footnote{Mike Shelley, Private communication, 2009}
\cite{ShelleyPRL07}.  Note that their definition
of $S_p^*$ includes the logarithmic correction from slender body theory
 and reads
$S_p^*=\frac{16\eta\dot{\epsilon}}{Y (\ln(L/R)-1/2)}\frac{L^4}{r^4}$. We correct the experimentally obtained value for the buckling threshold in the same way and find $S_p^* \sim$400. 
In the experiments the fibers are only half-immersed and viscous forces are certainly smaller then those used in the theoretical analysis. This is likely to be the main reason for the higher threshold
observed in our experiments.
Furthermore, even if we only consider filaments passing close to the stagnation point, 
the filament is of significant length compared to the typical lengthscale $W$ of the
cellular flow and the
approximation of hyperbolic streamlines does not hold in our experiments. 

We do not observe an abrupt increase of the
probability towards 100$\%$ at the threshold but a continuous
increase with $S_p$. At the highest value of $S_p$ accessible in our
experiments the filament still only buckles one out of four times it
passes a stagnation point. This indicates that even if the
threshold of buckling is controlled by $S_p$ the dynamics in the cellular flow are more complicated.  It
is likely that the probability might increase further when going to
higher values of $S_p$.

\begin{figure}[ht!]
\begin{center}
\includegraphics[width=7.8cm]{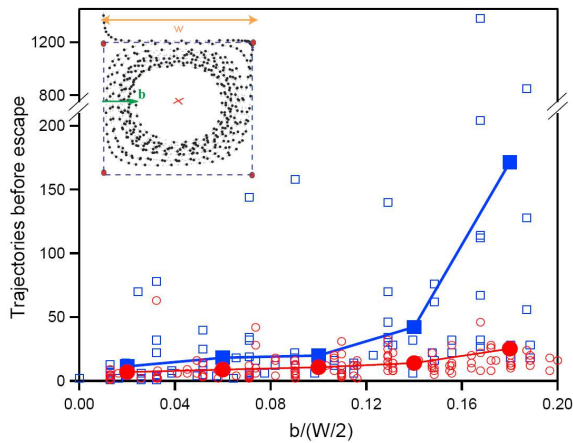}
\caption{\label{fig:diff} Number of
trajectories before escape as a function of $b/(W/2)$. Open symbols, individual experiments; closed symbols, the mean value. Red circles $S_p>120$, blue squares $S_p<120$.  Inset, definition of b.}
\end{center}
\end{figure}

With our experiment we can also address an other fundamental
question which stands in the modification of the translational
dynamics of the filament induced by its deformation. To do so we study
the escape of a filament from a vortex.
The finite size of the fiber already changes its dynamics in
the cellular flow compared to a pointlike particle (see figure{\ref{fig1}). We will thus
compare deformable fibers ($S_p>120$) to rigid fibers ($S_p<120$). The
fibers are placed in a vortex parallel to
the axis of compression and at a given distance $b$ (see figure \ref{fig:diff}c). We then record the number of trajectories
the filament needs to escape from the vortex and the number of
buckling events during this escape for over 250 experiments.

Figure \ref{fig:diff} shows the number of trajectories as well as
their average above and below the buckling threshold $S_p=120$ as a
function of $b$ normalized by the size of the vortex $W/2$. Note that the range of $b$ studied
corresponds to trajectories of the filament reaching at least 80~$\%$
of the maximum compression rate, identical to the experiments used
to study the buckling probability.

Trivially, one observes that the filament needs longer to escape
from a vortex if it is further away from the axis of compression.
One observes also that even the rigid filaments escape from the
vortices after a finite number of trajectories, showing the
importance of the effect of the finite size of the filament. The number of trajectories needed to
escape from a vortex fluctuates strongly in the case of the rigid
fibres, indicating that small changes in the initial position (not
measured in this analysis) are important for the dynamics. Less
fluctuations are observed for the flexible filaments. When comparing
the average number of trajectories of the flexible fibers and the rigid fibers one observes that the flexible filaments escape significantly faster. This is a clear signature of the effect of the
filament's deformation on its transport properties.

Young and Shelley show that the onset of the
buckling instability is coupled to the onset of a
translational "Brownian like" motion of the fiber across the lattice
of stagnation points \cite{ShelleyPRL07}. Above
the stretch coil transition, fibers present a diffusive behavior,
with however, a non trivial dependence of the diffusion coefficient
with the relative intensity of viscous and elastic forces. The motion of
a fiber in one cell can be seen as a 'collision' with the stagnation
point. Sequences of these collisions can then lead to diffusion of a
fiber. Our results are the first indication that
flexible fibers indeed diffuse faster than rigid ones through the cellular flow.

In conclusion, we have built an experimental setup allowing to study
the deformation and the transport of a centimetric fiber in a
cellular flow, formed by counter-rotating vortices. We have shown
that the elastic fiber can buckle due to the compressive viscous
forces. We have identified the threshold for the buckling
instability, which is function of the so called $S_p$ number, being a
measure of the relative intensity of viscous and elastic forces. These
experimental observations are in reasonable agreement with
linear stability analysis in a purely hyperbolic flow. We have also studied the translational
dynamics of the filament in ºthe cellular flow field. We have shown experimentally that fibers can
escape from a given vortex due to their finite size compared to the
size of the vortex. Flexible fibers are shown to escape after
significantly less tours compared to rigid fibers. This is a first
evidence that the deformation of a filament changes its transport
properties in the cellular flow.

We thank Mike Shelley for having initiated and closely followed this
study. We have benefitted from many enlightening discussions with
him. We thank Yuan-Nan Young, Olivier Cardoso and Denis Bartolo for fruitful
discussions. We acknowledge Jose Lanuza for supporting us in the
construction of the set-up and Guylaine Ducouret for help with the
rheological measurements.

\bibliography{buckledintranslation} %your .bib file
\bibliographystyle{unsrt}
\end{document}